\documentclass[manuscript]{acmart}
\usepackage{float}
\usepackage{longtable}
\usepackage{graphicx}
\usepackage{caption}
\usepackage{makecell}
\usepackage{xcolor}
\usepackage{subcaption}
\usepackage{longtable}
\usepackage{array}
\usepackage{multirow}
\usepackage{comment}
\usepackage{tabularx}  
\usepackage{ragged2e}  
\usepackage{booktabs}  
\usepackage{array}     
\usepackage{booktabs}   
\usepackage{rotating}   
\usepackage{geometry}   
\usepackage{color}
\usepackage[skip=2pt]{caption}
\excludecomment{appendixcontent}
\AtBeginDocument{%
  }
\usepackage{lipsum} 

\begin{document}
\title{CO-OPERA: A Human-AI Collaborative Playwriting Tool to Support Creative Storytelling for Interdisciplinary Drama Education}

\author{Xuejiao Ma}
\email{mxjinb612@gmail.com}
\orcid{0009-0006-6225-0445}
\authornotemark[1]
\affiliation{%
  \institution{East China Normal University}
  \city{Shanghai}
  \country{China}
}

\author{Haibo Zhao}
\email{hzhaobr@connect.ust.hk}
\orcid{0009-0004-7299-6879}
\authornotemark[2]
\affiliation{%
  \institution{Hong Kong University of Science and Technology}
  \city{Hong Kong}
  \country{Hong Kong}
}

\author{Zinuo Guo}
\email{51275901001@ecnu.edu.cn}
\orcid{0009-0006-6225-0445}
\authornotemark[3]
\affiliation{%
  \institution{East China Normal University}
  \city{Shanghai}
  \country{China}
}

\author{Yijie Guo}
\email{bjiang@deit.ecnu.edu.cn}
\orcid{0000-0001-9887-8548}
\authornotemark[4]
\affiliation{%
  \institution{The Future Laboratory, Tsinghua University}
  \city{Beijing}
  \country{China}
}

\author{Guanhong Liu}
\email{liugh@tongji.edu.cn}
\orcid{0000-0002-3266-2178}
\authornote{Corresponding author}
\affiliation{%
  \institution{Intelligent Big Data Visulization Lab}
  \institution{College of Design and Innovation,Tongji University}
  \city{Shanghai}
  \country{China}
}

\author{Bo Jiang}
\email{bjiang@deit.ecnu.edu.cn}
\orcid{0000-0002-7914-1978}
\authornote{Corresponding author}
\affiliation{%
  \institution{Lab of Artificial Intelligence for Education, East China Normal University}
  \city{Shanghai}
  \country{China}
}

\begin{abstract}
Drama-in-education is an interdisciplinary instructional approach that integrates subjects such as language, history, and psychology. Its core component is playwriting. Based on need-finding interviews of 13 teachers, we found that current general-purpose AI tools cannot effectively assist teachers and students during playwriting. Therefore, we propose CO-OPERA - a collaborative playwriting tool integrating generative artificial intelligence capabilities. In CO-OPERA, users can both expand their thinking through discussions with a tutor and converge their thinking by operating agents to generate script elements. Additionally, the system allows for iterative modifications and regenerations based on user requirements. A system usability test conducted with middle school students shows that our CO-OPERA helps users focus on whole logical narrative development during playwriting. Our playwriting examples and raw data for qualitative and quantitative analysis are available at \textit{https://github.com/daisyinb612/CO-OPERA}.
\end{abstract}
\maketitle
\vspace{-20pt}
\begin{figure}[H] 
    \centering 
    \includegraphics[width=1\textwidth]{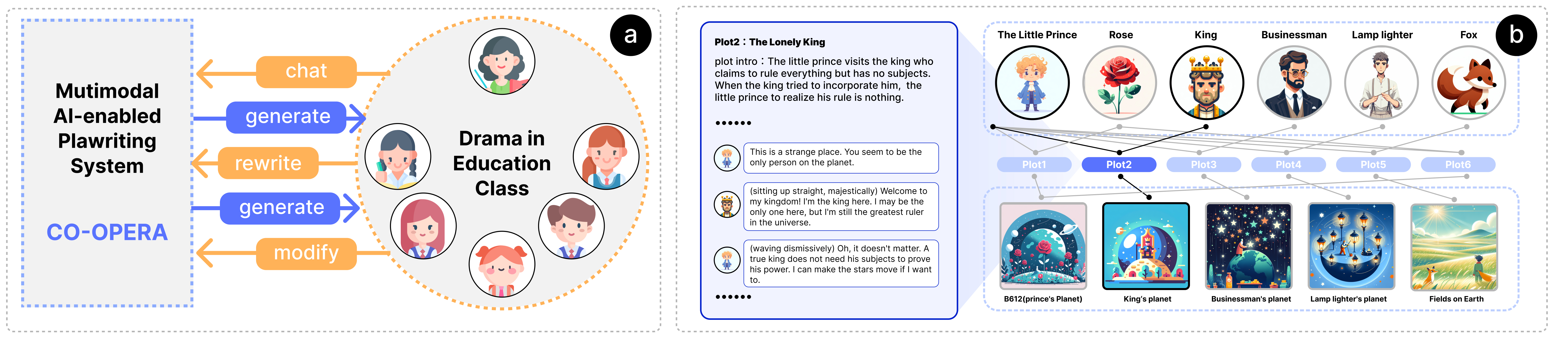} 
    \caption{ Overview: (a) Human-AI co-creation framework; (b) Playwriting examples with dramatic elements  
}  
\end{figure}
\vspace{-20pt}
\section{Introduction}
\textbf{Drama-in-education} is an instructional approach that uses experience simulation and role-playing \cite{liuWhenHeFeels2024,guneysuEnhancingAutismTherapy2024} to cultivate students’ comprehensive qualities, such as creativity, collaboration and so on. Its main objective is to encourage learners to discover and create through active participation\cite{boothHowTheatreEducates2003}. \textbf{Playwriting}, as its core part, is essentially a creative cognitive activity that combines dynamic knowledge construction with narrative expression. 
Unlike other narrative genres, its plots are only driven by direct dialogue between characters without descriptive depiction. 

For playwriting, students face \textbf{certain obstacles} including lack of overall narrative perspective, ignoring the causes of character personality etc. 
These obstacles hinder students' sustained engagement in consistent creative and critical cognitive activity. 
For instance, history teachers may ask students to create scripts to deepen their understanding of historical events. 
Specifically, students often focus excessively on crafting literal dialogues while overlooking the dialectical analysis of historical narratives.Even though teachers want to integrate generative AI into playwriting instruction, current AI tools lack scaffolding mechanisms for students to think systematically. Therefore, designing a dedicated playwriting system for drama-in-education classes is necessary.  

Recently, researchers have focused on how AI can support the creation of effective storytelling tools \cite{bernalPaperDreamsRealTime,hanDesignImplicationsGenerative2023,zhangStoryDrawerChildAI2022,zhangStoryBuddyHumanAICollaborative2022}. Unlike traditional systems, AI-driven tools provide the generation of story content and materials \cite{hanDesignImplicationsGenerative2023,zhangStoryDrawerChildAI2022}. However, most of the research focuses on supporting storytelling for children and families \cite{zhangStoryBuddyHumanAICollaborative2022,dietzStoryCoderTeachingComputational2021a,fanStoryPromptExploringDesign2024,panjwaniConstructingMeaningDesigning2017,chen2025storymate,10.1145/3706598.3713602,10.1145/3706598.3713478}, such as language learning \cite{koryStorytellingRobotsLearning2014,druinDesigningIntergenerationalMobile2009}, emotional expression learning \cite{ryokaiStoryFacesPretendplayEbooks2012,schlauchInvestigatingSocialEmotional2022}, mathematics learning \cite{alexandreMaths4KidsTellingStories2010,xuMathKingdomTeachingChildren2023,zhangMathemythsLeveragingLarge2024a}, and computational thinking and programming \cite{dietzStoryCoderTeachingComputational2021a,chenChatScratchAIAugmentedSystem2024}. Among them, relatively few studies involve how to support educational playwriting. 

Even some previous research has explored leveraging language models for story generation and plot coherence control. Notably, Yao et al \cite{yaoPlanWriteBetterAutomatic2019c}. proposed a hierarchical story generation strategy to enhance narrative consistency. Futhermore, there is \textit{Dramatron} \cite{mirowskiCoWritingScreenplaysTheatre2023}, a script creation tool for professional screenwriters, and IBSEN \cite{hanIBSENDirectorActorAgent2024,wuRolePlayDramaInteractionLLM2024b} uses only AI to generate whole scripts, their interactive methods and script themes are not suitable for educational purposes.

Our semi-structured interviews was conducted with interdisciplinary teachers (N=13) using playwriting teaching methods, we identified \textbf{two core challenges}: (1) Students' fragmented creative ideas in drama classes \textbf{lack structural integration}, making it difficult to synthesize them into coherent scripts; (2) General \textbf{AI tools fail to satisfy the demand} for realistic educationally complex narratives in interdisciplinary drama education. Thus, we aim to explore the design space of human-AI interaction playwriting to support teachers and students create complex plots that fit different educational purposes.

Therefore, we introduce CO-OPERA, a collaborative creation platform for educational playwriting that integrates generative AI. On this platform, educators and learners collaboratively leverage generative AI and intelligent tutoring to co-create instructional playwriting outlines that can be rapidly deployed for structured pedagogical discussions. Based on the two cognitive process of creation, the system constructs a \textbf{“divergence-convergence” interaction} mechanism: promoting divergence thinking through conversational guidance by intelligent tutors, and achieving convergence thinking by the generation from a multi-agent framework. 

 This study presents the following two main contributions: 
 \begin{itemize}
    \item We identified two core challenges in educational scriptwriting and current AI tool limitations through formative studies, including needs-finding interviews with teachers (N=13) using playwriting as an instructional method.
    \item We designed and developed CO-OPERA, a human-AI collaborative playwriting system, to support educational classrooms. Its usability was preliminarily validated through a System Usability Scale (SUS) evaluation with students (N=12) in a middle school setting.
\end{itemize}
\section{Formative Study}
\subsection{Participants}
Semi-structured interviews with 13 participants was conducted. The participants were recruited through a contact list from our previous study and public recruitment on \textit{rednote}, a social media platform. The demographics of 13 participants is shown in Table 1. 

\vspace{-20pt}
\begin{table}[H]
\centering
\footnotesize
\label{tab:participants}
\caption{Summary of study participants}
\renewcommand{\arraystretch}{1}
\begin{tabularx}{\linewidth}{ 
  >{\RaggedRight}p{0.5cm}  
  >{\Centering}p{0.7cm}  
  >{\Centering}p{1.5cm}  
  >{\Centering}p{4cm}  
  >{\Centering}p{3.5cm}  
  >{\Centering}p{1.2cm}  
  >{\Centering}p{0.5cm}  
}
\toprule
\textbf{No} & \textbf{Gender} & \textbf{Major} & \textbf{Teaching Identity} & \textbf{Education Goals} & \textbf{Age Range} & \textbf{EXP(Year)} \\
\midrule
P1 & F & Preschool Ed. & head of Drama Institution & Communication, creativity & 4-12 & 9 \\
P2 & F & Philosophy & drama club teacher & Philosophical thinking & 12-17 & 5 \\
P3 & F & Drama Ed. & Drama/Chinese teacher & Critical thinking, moral education & 6-12 & 3 \\
P4 & F & Drama Ed. & Public elementary drama teacher & Imagination, empathy & 6-12 & 7 \\
P5 & F & English & Kindergarten teacher & English, teamwork skills & 4-6 & 4 \\
P6 & M & Sociology & Head of Drama Institution & Happiness, English & 3-12 & 10 \\
P7 & M & Directing & Chinese writing instructor & Writing skills & 10-12 & 2 \\
P8 & M & Performance & Head of performing arts school & Performance theory & 15-17 & 10 \\
P9 & F & Preschool Ed. & Public kindergarten teacher & Communication & 3-4 & 2 \\
P10 & F & Pedagogy & Int'l primary drama teacher & English proficiency & 5-12 & 4 \\
P11 & M & Drama Ed. & Head of Drama Institution & Problem solving & 4-9 & 10 \\
P12 & M & Performance & Independent tutor & Critical thinking & 6-12 & 10 \\
P13 & F & Drama Ed. & PT middle school drama club & Independent thinking & 12-14 & 5 \\
\bottomrule
\end{tabularx}
\end{table}
\vspace{-30pt}
\subsection{Procedure}
The researchers conducted semi-structured interviews with participants via \textit{Tencent Meeting}, a online meeting platform, with each session lasting between 30 and 60 minutes. Participants received monetary compensation for their involvement. Following data collection, the recorded interviews were transcribed verbatim, anonymized and imported into Nvivo software for qualitative analysis.

\subsection{Findings}
Generally speaking, current pedagogical practices in playwriting development demonstrate two primary approaches: (1) the adaptation of canonical narratives (P3, P4, P6, P10) and (2) innovative adaptation processes, which have two stages: narrative framework(storyline and plots) and detailed dialogue. They are mainly developed through four methods: (a) Educators setting a narrative framework (P3, P6); (b) Educators revising scripts based on students’ ideas (P3, P6, P11, P13); (c) Students creating scripts according to the framework (P6, P10); (d) Student self-driven writing with educators’ help only when needed (P1-P5, P7, P8, P12).

Two challenges are summarized: (1) Challenge 1: There is a lack of playwriting logic scaffolding in class to assist students in script creation, as introduced in 3.2.1; (2) Challenge 2: The current general AI tools fall short in meeting the creative demands for crafting complex plays, as introduced in 3.2.2.

\subsubsection{\textbf{Barriers in the current class}}
Our interview has found that students’ lack of life experiences impede the generation of inspiration, while their introverted personalities impede the expression of creative ideas:
P8:\textit{"Students who don’t have much life experience—when they try to create stuff, there are things they just can’t come up with, or even imagine."};
P4:\textit{"Like, a teacher might give a topic and think there’s tons to explore, but the student? They’re stuck trying to find inspiration from it."};
P7:\textit{"Some kids are naturally quiet, or maybe they’re new to the class and still awkward. That’s when teachers gotta nudge them gently and cheer them up, you know?"}

Moreover, students who focus on refining specific sentences in the early stages  may demonstrate insufficient awareness of overall thinking and compositional logic.
P7:\textit{"Young students often fall into this writing trap—they get super hung up on tweaking one or two specific dialogues. That’s when I jump in and nudge them to see the bigger picture."}


\subsubsection{\textbf{Current limitations of AI tool for playwriting}}
We further investigated the limitations of current general-purpose AI tools in playwriting tasks, which can be categorized into three levels.
(1) High inter- and communication costs:
P5:\textit{"You’ve got to spell everything out for the AI—how many characters, what exactly you want. AI often misunderstands me.After wasting hours tweaking pro, I could’ve written this faster myself!So frustrating."}
P8:\textit{"AI just doesn’t create like humans. Even if I feed it specifics—characters, time, place—it still can’t grasp what I’m really going for. Night and day difference."}
; (2) Naively simplistic and monotonous outputs:
P3:\textit{"If you just give the AI a topic to write, it’ll probably crank out childish stuff."}
P12:\textit{"AI’s stories always feel recycled—maybe because the data it’s fed is just the same generic stuff everyone’s used."}
; (3) Lack of deep comprehension and innovative capabilities:
P6:\textit{"You can’t just code human relationship into numbers—so the AI totally misses how it actually shape character growth."}
P4:\textit{"If I’m hunting for fresh ideas? Tough luck. AI’s stuck recycling the same old data patterns—zero real innovation."}

\section{CO-OPERA}
\subsection{System Overview}
CO-OPERA is powered by agents and tutors based on a multimodal LLM. As shown in Figure 2, teachers and students can interact with the system according to the following steps:
\textbf{(a) Input logline:} Users can input a brief summary of their drama story.
\textbf{(b) Character:} Users can automatically generate character information.
\textbf{(c) Plot:} Users can automatically generate plot elements, allowing them to modify specific details.
\textbf{(d) Scene:} Users can automatically generate scene information.
\textbf{(e) Dialogue:} Users can generate and modify dialogues.
\vspace{-10pt}
\begin{figure}[h]
    \centering
    \includegraphics[width=1\linewidth]{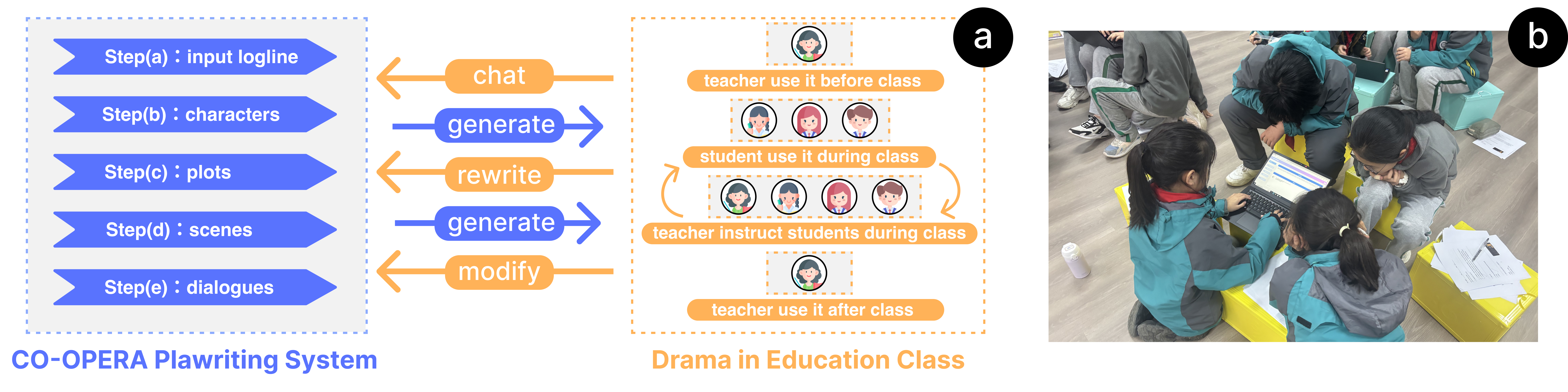}
    \caption{(a)Human-AI interaction framework of playwriting; (b) Interdisciplinary drama class through teamwork}
\end{figure}
\vspace{-20pt}
\subsection{Multi-agent Workfloww}
For each step, we employ two types of agents that guide users in playwriting creation. The \textbf{functional agent} generates standardized elements of playwriting, while the \textbf{conversational tutor} engages users through dialogic coaching to broaden their creative thinking and clarify their artistic vision.As shown in Figure 3, taking step (a) as an example, the user can refine the story summary with the \textbf{'summary tutor'} by asking and answering. Once satisfied, the user can submit their logline to the logline box and click the confirm button to upload the \textcolor{red}{logline} data. This data is then input into the \textbf{'character agent'} for \textcolor{orange}{character list} data generation. In this way, we gradually discuss and generate \textcolor{green}{plots}, \textcolor{cyan}{scene}, and \textcolor{blue}{dialogue} data, and finally get a complete script.

\textbf{System Implementation}:The frontend interactive web application of CO-OPERA is built using the Vue3 framework, while the backend is developed with Python using the Flask framework. Text and image functionalities all utilize the GPT-4o API. 
\vspace{-10pt}
\begin{figure}[H]
    \centering
    \includegraphics[width=0.9\linewidth]{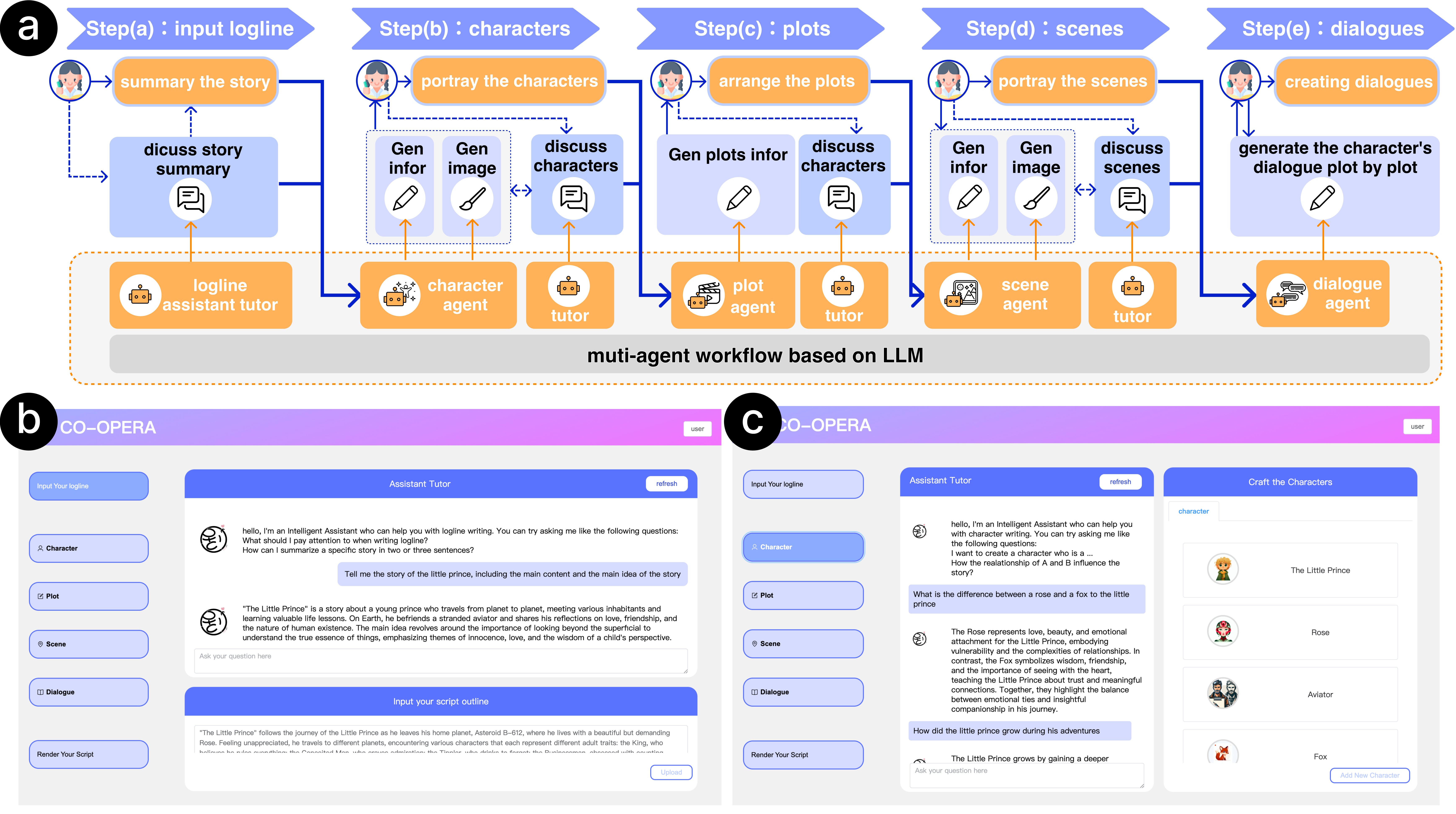}
    \caption{Multi-agent workflow: dataflow among functional agents, assistant tutors, and users.}
\end{figure}
\vspace{-10pt}

\vspace{-10pt}
\section{Evaluation}
\subsection{The editing of playwriting elements reflects users' creation}
\textbf{Participant and Process:}We participated in a high school drama class where students initiated collaborative brainstorming. Using the CO-OPERA system, they created logline and characters for 30 mins.One week later, they created plots and dialogues for another 30 mins . We collected data comparing both the original AI-generated vision and their final version.

\textbf{Result:}Figure 4(a) shows the absolute edit distance after normalization, which represents the difference in text length. Figure 4(b) compares the two text versions, including the relative edit distance (deletion length and insertion length) and Jaccard similarity. 
These revisions reflect CO-OPERA's capacity to help users complete discussions and critical thinking throughout playwriting development.

\subsection{The application with a questionnaire reflects user feedback}
\textbf{Participant and Process:}We were involved in a middle school psychological drama class. Five groups of students used CO-OPERA for about 30 minutes to create their scripts according to the teacher's theme. The two or three students in each group operating the system completed SUS questionnaires, resulting in 12 valid responses, which aligns with the recommended range.In the questionnaire, items Q1, 3, 5, 7, and 9 were positively worded, whereas Q2, 4, 6, 8, and 10 were negatively worded. All responses were reverse-scored according to SUS guidelines for statistical analysis.

\textbf{Result:}This study obtained a comprehensive score of 77.08 (SD=13.35) using the System Usability Scale (SUS). According to the benchmark study by Bangor et al. (2009)\cite{bangorDeterminingWhatIndividual2009}, the usability of the system is good(higher than the industry benchmark of 68 points).
As shown in Figure 4(c), the system demonstrates strong performance in Functional Cohesion (M=3.29), aligning with participant feedback that it \textit{"enhances logical structuring for playwriting."} However, it underperforms in Practical Usability (M=2.83), which is consistent with critical user remarks regarding \textit{"unacceptably slow generation speeds."}
Although substantial opportunities for refinement remain, experimental results confirm CO-OPERA's capability to deliver logic-driven support for playwriting.
\vspace{-12pt}
\begin{figure}[H]
    \centering
    \includegraphics[width=1\linewidth]{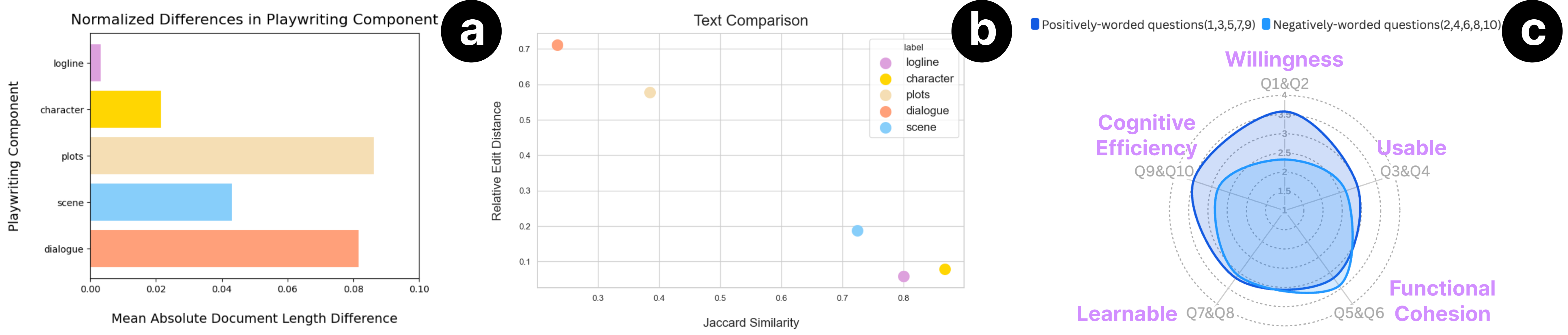}
    \caption{(a) Absolute edit distance; (b) Relative edit distance; (c) SUS questionnaire results:Willingness (Mean=3.04; Q1=3.58, Q2=2.50), Usable (Mean=3.04; Q3=3.58, Q4=2.50),Functional Cohesion (Mean=3.29; Q5=3.17, Q6=3.42), Learnable (Mean=3.13; Q7=3.17, Q8=3.08), Cognitive Efficiency (Mean=3.13; Q9=3.42, Q10=2.83).}
\end{figure}
\vspace{-20pt}

\vspace{-5pt}
\section{Discussion}
\vspace{-5pt}
The results from our user study suggest that users can successfully collaborate with CO-OPERA to create complex interactive playwriting experiences for educational purposes. CO-OPERA also performed well in helping students construct creative story ideas and to further think about and modify AI-generated content.Below, we discuss the implications and design themes emerged from our work.

\textbf{Playwriting creation an interdisciplinary instructional  approach:} Through interviews with teachers, we gained a deeper understanding of how playwriting is utilized in different interdisciplinary classrooms. Unlike in subject teaching, teachers not only focus on the quality of the final output but also emphasize students’ discussions and reflections on real life through creative processes, aligning with the theory of ‘learning by doing’\cite{anzai1979theory} and ‘Situated Learning’\cite{lave1991situated} , which foreground knowledge construction through authentic social participation. Additionally, during the creative process\cite{davis2013toward}, ‘conversational tutor’ and ‘drama-element agent’ have been implemented to support human cognitive mechanisms, such as divergent thinking\cite{Guo2020Tangible,Lin2020Collaborative} and convergent thinking\cite{Jeon2021FashionQ}. 

\textbf{Role of AI in eductional playwriting:} The design of CO-OPERA introduced generative AI into an existing activity that previously involved only teachers and students. Should the AI system be a peer, an agent of the teacher, a tool, or something else? In our research, the "conversational tutor" helps students think divergently and enrich their inspiration, while the "drama-element agent" gradually generates structured content based on students’ inputs and modifications. Since the course goal is dialectical thinking, generative AI can provide students with multiple perspectives and creative materials. Unlike general AI tools that deliver results in one step, CO-OPERA adopts step-by-step generation, offering teachers and students discussable content at each phase, such as themes, character personalities, and plot causality. 

\section{Conclusion}
\vspace{-5pt}
Although the findings are promising, this study has some limitations. The sample size was limited. Our future work should involve collaboration with additional drama courses and controlled experiments to explore multi-modal AI's educational support methods for playwriting.

Our study first conducted need-finding interviews with 13 practicing drama teachers to identify design requirements. Based on these insights, we developed the CO-OPERA system to support playwriting in drama education.Subsequently, we recruited 12 high school students for SUS test. The qualitative and quantitative results indicate that our method has basically responded to our challenges.


\bibliographystyle{ACM-Reference-Format}
\bibliography{references}

\appendix
\section{Research Materials}
\subsection{Semi-structured Interviews}
\subsubsection{Teaching Background and Personal Experience}
\begin{enumerate}
    \item \textit{May I ask where you teach and which subjects you teach?(e.g.private or public schools/Chinese, English or Drama)}
    \item \textit{What initially motivated you to incorporate playwriting into your teaching?}
    \item \textit{How many years of teaching experience do you have?}
    \item \textit{Who are your primary students? (e.g., age groups, educational backgrounds)}
\end{enumerate}

\subsubsection{Teaching Objectives, Structure, and Methods}
\begin{enumerate}
    \item \textit{In your opinion, what are the most significant values of drama in education?}
    \item \textit{Different from other educational approaches, what specific skills do you think it helps students develop?}
    \item \textit{How frequently are classes held? (e.g., weekly or sessions) How do you plan your teaching content for each class?}
    \item \textit{How do you incorporate playwriting in your teaching?}
    \item \textit{What teaching methods do you employ to help students in playwriting? (e.g., group activities, role-playing)}
    \item \textit{On average, how many students are in one class?}
    \item \textit{How many groups are formed in one class? How many students are in each group?}
    \item \textit{How are roles divided for students during class?}
\end{enumerate}

\subsubsection{Teaching Objectives, Structure, and Methods}
\begin{enumerate}
    \item \textit{How do you integrate playwriting content into your classes?}
    \item \textit{What types of stories or scripts are commonly used? (e.g.fables, historical narratives, etc.)}
    \item \textit{Who leads the playwriting process? (e.g.students, teachers, or collaboration?) Could you provide an example?}
    \item \textit{What tools or methods are typically used in playwriting creation?}
    \item \textit{What qualities do you consider essential for effective educational playwriting?}
    \item \textit{What criteria do you use to evaluate students’ playwriting work?}
    \item \textit{How do students and parents generally perceive drama education?}
    \item \textit{What specific challenges have you encountered in playwriting? Could you provide an example?}
\end{enumerate}

\subsubsection{AI-Based Playwriting in Teaching}
\begin{enumerate}
    \item \textit{Have you integrated AI tools into playwriting activities? If yes, what benefits or challenges did you observe? Please share an example}
    \item \textit{In which specific stages of scriptwriting do you think AI systems could provide assistance?}
    \item \textit{How can AI tools help teachers or students in playwriting?}
    \item \textit{Are there any stages where you would avoid using AI? Why?}
    \item \textit{What possibilities do you envision for AI-enhanced scriptwriting in education?}
    \item \textit{Do you have any further insights or experiences related to drama education or AI-integrated playwriting that you would like to share?}
\end{enumerate}

\subsection{SUS questionnaire}
\begin{enumerate}
    \item I would be willing to use CO-OPERA for playwriting.
    \item I found CO-OPERA overly complicated to operate.
    \item I thought CO-OPERA was easy to use.
    \item I would need guidance from someone to use CO-OPERA.
    \item I found the various functions in CO-OPERA well-integrated.
    \item I thought there was inconsistency in how CO-OPERA's functions worked together.
    \item I believe most people could learn to use CO-OPERA very quickly.
    \item I found CO-OPERA very cumbersome to use.
    \item I felt very confident using CO-OPERA.
    \item I needed to learn a lot of things before I could use CO-OPERA.
\end{enumerate}

\end{document}